\documentclass[fleqn,twoside]{article}
\usepackage{espcrc2}
\usepackage{graphicx}
\usepackage{epsf}

\usepackage{amsmath, amssymb,latexsym,float,algorithm,algorithmic}
\usepackage{enumerate}

\newcommand{\Real}{\mathbb R}
\newcommand{\Complex}{\mathbb C}

\newcommand{\R}{\text{\fontshape{n}\selectfont I\kern-.42exR}}

\newcommand{\1}{\text{\fontshape{n}\selectfont 1\kern-.56exl}}

\title{Lanczos Methods for UV-Suppressed Fermions}

\author{A. Bori\c{c}i
       \address{University of Edinburgh, Department of Physics and Astronomy \\
                Mayfield Rd, JCMB, Edinburgh EH9 3JZ}
       }

\begin{document}

\begin{abstract}
In this talk I indroduce lattice fermions with suppressed
cutoff modes. Then I present Lanczos based methods for
stochastic evaluation of the fermion determinant.
\vspace{1pc}
\end{abstract}

\maketitle

\section{INTRODUCTION}

There has been recent interest in the so-called `Ultraviolet slowing
down' of fermionic simulations in lattice QCD
\cite{Irving_et_al,Duncan_et_al,PhdF,Peardon,AHasenfratz_Knechtli}.
These studies try to address algortmically large fluctuations
of the high end modes of the fermion determinant. The goal is to
increase the signal-to-noise ratio of the infrared modes and to
accelerate fermion simulations as well.

In fact, all the computational effort needed to treat UV-modes
by above algorithms can be reduced to zero by suppressing them
in the first place \cite{Borici_UVSFa}. The lattice Dirac
operator of this fermion theory is given by:
\begin{equation}\label{D_operator}
D = \frac{\mu}{a} \Gamma_5 \tanh \frac{a \Gamma_5 D_{W/S/O}}{\mu}
\end{equation}
where $D_{W/s/o}$ is the input lattice Dirac operator, $a$ the
lattice spacing and $\mu > 0$ is a dimensionless parameter.
For Wilson (W) and overlap (o) fermions as the input theory one has
$\Gamma_5 = \gamma_5$. For staggered fermions $\Gamma_5$ is a diagonal
matrix with entries $+1/-1$ on even/odd lattice sites.
The theory converges to the input theory in the contimuum limit
and is local and unitary as shown in detail in
\cite{Borici_UVSFa}. The input theory is also recovered in the limit
$\mu \rightarrow \infty$. For $\mu \rightarrow 0$ one has
$D \rightarrow \mu$, i.e. a quenched theory.

Perturbative calculations with this theory are straightforward.
In the following Wilson fermions are used as input.
The inverse fermion propagator is given by:
\begin{equation}\label{inverse_propagator}
{\tilde D}(p) = \frac{\mu}{a} \gamma_5 \tanh \frac{a {\tilde
H_W}(p)}{\mu}
\end{equation}
with $p = \{p_{\nu}, \nu=1,\ldots,4\}$ being the four-momentum
vector.
As usual, gauge fields are parametrized by $su(3)$
elements, $U(x)_{\nu} = e^{i a g A(x)_{\nu}}$ with
$A(x)_{\nu} \in su(3)$ algebra,
and the Wilson operator is written as a sum of
the free and interaction terms: $D_W = D_W^0 + D_W^I$.
The splitting of the lattice Dirac operator is written
in the same form:
\begin{equation*}
D = D^0 + D^I,
~~~~D^0 = \frac{\mu}{a} \gamma_5 \tanh \frac{a H_W^0}{\mu}
\end{equation*}
where the interaction term has to be determined.
This can be done by expanding $D$ in terms of $a/\mu$:
\begin{equation}\label{pert_expan}
D = D_W[\1 + c_1 (\frac{a H_W}{\mu})^2
           + c_2 (\frac{a H_W}{\mu})^4 + \cdots]
\end{equation}
where $c_1, c_2, \ldots$ are real expansion coefficients.
Calculation of $D^I$ is an easy task if one stays with a finite
number of terms in the right hand side of (\ref{pert_expan}).
Also, the number of terms can be minimized using a
Chebyshev approximation for the hyperbolic tangent.
\footnote{I would like to thank Joachim Hein for discussions
related to lattice perturbation theory.}

\section{COMPUTATIONAL METHODS}

The effective action of the theory defined above can be written as:
\begin{equation}\label{eff_action}
S_{\text{eff}} = \text{tr} f(A)
\end{equation}
where $A \in \Complex^{N\times N}$ and $f(s)$
is a real and smooth function of $s \in \Real^{+}$. The matrix $A$
is assumed to be Hermitian and positive definite.
Since the trace is dificult to obtain one can use the stochastic
method of \cite{Bai_et_al}. The method is based on evaluations of
many bilinear forms of the type:
\begin{equation}\label{b_forms}
{\mathcal F}(b,A) = b^T f(A) b
\end{equation}
where $b \in \Real^N$
is a random vector. The trace is estimated as an average over many
bilinear forms. A confidence interval can be computed as described
in detail in \cite{Bai_et_al}. Here I will
describe an alternative derivation of the Lanczos algorithm \cite{Lanczos},
which is used in \cite{Bai_et_al} and in \cite{Cahill_et_al} as well.
The full details of this study can be found at \cite{Borici_UVSFb}.
It uses familiar tools (in lattice simulations) such as
sparse matrix invertions and Pad\'e approximats.

The Pad\'e approximation
of a smooth and bounded
function $f(.)$ in an interval
can be expressed as a partial fraction expansion:
\begin{equation}
f(s) \approx \sum_{k=1}^m \frac{c_k}{s + d_k}
\end{equation}
with $c_k \in \Real, d_k \geq 0, k = 1, \ldots, m$.
It is assumed that the right hand side converges to the left hand side
when the number of partial fractions becomes large enough.
For the bilinear form I obtain:
\begin{equation}\label{partial_frac}
{\mathcal F}(b,A) \approx \sum_{k=1}^m b^T \frac{c_k}{A + d_k \1} b
\end{equation}
A first algorithm may be derived at this point.
Having computed the
partial fraction coefficients one can use a multi-shift iterative
solver of \cite{Freund} to evaluate the right hand side (\ref{partial_frac}).
The problem with
this method is that one needs to store a
large number of vectors that is proportional
to $m$. This could be prohibitive if $m$ is say larger than $10$.
However, it is easy to save memory if only the bilinear form
(\ref{b_forms}) is needed (see \cite{Borici_UVSFb} for details).

If a Pad\'e approximation is not sufficient or difficult to obtain,
the Lanczos method is the only alternative
to evaluate exactly the bilinear forms of type
(\ref{b_forms}).
To see how this is realized I assume that the linear system $A x = b$
is solved to the desired accuracy using the Lanczos algorithm.
The algorithm produces the coefficients of the Lanczos matrix $T_n$,
which is symmetric and tridiagonal.
Its eigenvalues, the so called Ritz values, tend to approximate the
extreme eigenvalues of the original matrix $A$.
In the application considered here one can show that \cite{Borici_UVSFb}:
\begin{equation}\label{lemma}
\sum_{k=1}^m b^T \frac{c_k}{A + d_k \1} b = ||b||^2 \sum_{k=1}^m
e_1^T \frac{c_k}{T_n + d_k {\1}_n} e_1
\end{equation}
where ${\1}_n \in \Real^{n\times n}$ is the identity matrix and $e_1$
is its first column.
From this result and the convergence of the partial fractions to the
matrix function $f(.)$, it is clear that:
\begin{equation}\label{reduced_form}
{\mathcal F}(b,A) \approx
{\mathcal {\hat F}}_n(b,A) = ||b||^2 e_1^T f(T_n) e_1
\end{equation}
Note that the evaluation of the right hand side
is a much easier task than
the evaluation of the right hand side of (\ref{b_forms}).
A straightforward method is the spectral decomposition of the
symmetric and tridiagonal matrix $T_n = Z_n \Omega_n Z_n^T$,
where $\Omega_n \in \Real^{n\times n}$ is a diagonal matrix of
eigenvalues $\omega_1,\ldots,\omega_n$ of $T_n$ and $Z_n \in \Real^{n\times n}$
is the corrsponding matrix of eigenvectors, i.e. $Z_n = [z_1,\ldots,z_n]$.
From (\ref{reduced_form}) and the spectral form $T_n = Z_n \Omega_n Z_n^T$
it is easy to show that:
\begin{equation}\label{omega_form}
{\mathcal {\hat F}}_n(b,A) = ||b||^2 e_1^T Z_n f(\Omega_n) Z_n^T e_1
\end{equation}
where the function $f(.)$ is now evaluated at individual eigenvalues of
the tridiagonal matrix $T_n$.
The eigenvalues and eigenvectors of a symmetric and tridiagonal matrix
can be computed by the QL method with implicit shifts
\cite{Numerical_Recipes}. The method has an $O(n^3)$ complexity.
Fortunately, one can compute (\ref{omega_form}) with
only an $O(n^2)$ complexity.
Closer inspection of eq. (\ref{omega_form}) shows that besides the
eigenvalues, only the first elements of the eigenvectors are needed:
\begin{equation}\label{result_form}
{\mathcal {\hat F}}_n(b,A) = ||b||^2 \sum_{i=1}^n z_{1i}^2 f(\omega_i)
\end{equation}
It is easy to see that the QL method delivers the eigenvalues and
first elements of the eigenvectors with $O(n^2)$ complexity.
\footnote{I thank Alan Irving for the related comment on the QL implementation
in \cite{Numerical_Recipes}.}
A similar formula (\ref{result_form}) is suggested
by \cite{Bai_et_al} based on
quadrature rules and Lanczos polynomials.
\begin{algorithm}[htp]
\caption{Lanczos algorithm to compute (\ref{b_forms}).}
\label{lambda_algor}
\begin{algorithmic}
\STATE Set $\beta_0 = 0, ~\rho_1 = 1 / ||b||_2, ~q_0 = o, ~q_1 = \rho_1 b$
\FOR{$~i = 1, \ldots$}
    \STATE $v = A q_i$
    \STATE $\alpha_i = q_i^{\dag} v$
    \STATE $v := v - q_i \alpha_i - q_{i-1} \beta_{i-1}$
    \STATE $\beta_i = ||v||_2$
    \STATE $q_{i+1} = v / \beta_i$
    \STATE $\rho_{i+1} = - (\rho_i \alpha_i + \rho_{i-1} \beta_{i-1}) / \beta_i$
    \IF{$1 / |\rho_{i+1}| < \epsilon$}
       \STATE $n = i$
       \STATE stop
    \ENDIF
\ENDFOR
\STATE Set $~(T_n)_{i,i} = \alpha_i, ~(T_n)_{i+1,i} = (T_n)_{i,i+1} = \beta_i$,
       otherwise $~(T_n)_{i,j} = 0$
\STATE Compute $\omega_i$ and $z_{1i}$ by the QL method
\STATE Evaluate (\ref{b_forms}) using (\ref{result_form})
\end{algorithmic}
\end{algorithm}
Clearly, the Lanczos algorithm, Algorithm \ref{lambda_algor} has
$O(nN)+O(n^2)$ complexity, but it
delivers an exact evaluation of (\ref{b_forms}) and for typical applications
in lattice QCD the $O(n/N)$ overhead is small.
The method of
\cite{Bai_et_al} computes the relative differences of (\ref{result_form})
between two successive Lanczos steps and stops if they don't decrease
below a given accuracy. In order to perform the test their algorithm
needs to compute the eigenvalues of $T_i$ at each Lanczos step $i$.
This is illustrated in Fig. 1 where the convergence of the
bilinear form (\ref{b_forms}) and extreme eigenvalues of $A$ are plotted.
\begin{figure}
\epsfxsize=3.8cm
\vspace{2cm}
\hspace{2.5cm} \epsffile[240 400 480 450]{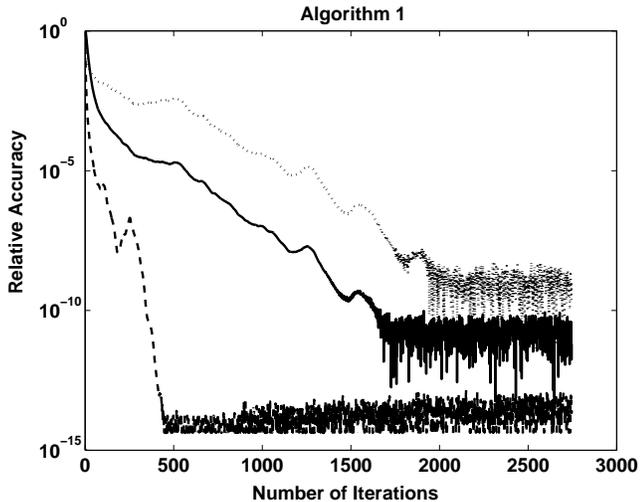}
\vspace{2.7cm}
\caption{Relative differences of (\ref{result_form}) (solid line),
the largest (dashed line) and the smallest eigenvalue (dotted line) of $T_n$
for $f(s) = log (tanh \sqrt{s}), s > 0$, $A = H_W^2$,
and $b \in \mathbb{Z}_2$ noise on a $12^3\times24$ lattice at
$\beta = 5.9$ and bare Wilson quark mass $m = -0.869$.}
\end{figure}
The figure sugests that one can compute the bilinear form less
frequently than proposed by \cite{Bai_et_al}.

\vspace{0.3cm}
\noindent
{\bf Acknowledgements}. I would like to thank
Philippe de Forcrand, Alan Irving and Tony Kennedy
for discussions related to this talk.

\end{document}